\documentstyle[prl,aps,epsfig,twocolumn]{revtex}
\newlength{\figwidth}
\begin{document}

\title{ Superscars}
\author{ E. Bogomolny and C. Schmit}
\address{Laboratoire de Physique Th\'eorique et Mod\`eles Statistiques\dag\\ 
 Universit\'e de Paris-Sud, B\^at. 100, 91405 Orsay Cedex, France}
\date{\today}
\maketitle

\begin{abstract}
Wave functions of plane polygonal billiards are investigated. It is
demonstrated that they have clear structures (superscars) related with
families of classical periodic orbits which do not disappear at large energy.

\vspace{2ex}
\noindent
PACS numbers: 05.45.Mt, 03.65.Sq, 42.25.Fx 
\end{abstract}

\vspace{3ex}

A central problem of quantum chaos is an adequate description of different
types of quantum systems which do not permit exact solution. For example,
it is well accepted that eigenenergies of  chaotic systems are distributed 
as eigenvalues of random matrix ensembles \cite{BGS} and their eigenfunctions
are described by Gaussian random functions \cite{Berry}. 
But much less is known when a model is neither chaotic or integrable. 

Particular intriguing examples are plane polygonal
billiards (PPB) whose classical mechanics is surprisingly rich.
When all their angles are rational with $\pi$ these models are called 
pseudo-integrable (PI) because their classical trajectories cover
two-dimensional surfaces of genus $g>1$ (see e.g. \cite{Richens}).
It was established numerically \cite{Gerland} that spectral
statistics of PI models  in many aspects resembles statistics of  
the Anderson model at the metal-insulator transition \cite{Shklovskii}. 
In particular, the nearest-neighbor distribution displays a linear level 
repulsion at small distances and an exponential decay at large separations. 

The purpose of this letter is to investigate wave functions
of certain PI systems. It is found that they have
superscaring property, namely, many of them have clear structures
connected with families of classical periodic orbits which
do not disappear at large energy. 

The scar phenomenon in chaotic systems is not new.
The existence of structures  near unstable periodic orbits in chaotic 
wave functions was established in \cite{Heller1} and later 
many works were done to clarify the subject (see e.g. \cite{Heller2} 
and references therein). 
PI systems differ in many aspects from chaotic and integrable
systems and to the authors knowledge no conjecture about their wave functions 
exists in the literature. 

The main difficulty with analytical treatment of PI models is 
the strong diffraction on billiard corners with angles $\neq \pi/n$ with 
integer $n$. When a bunch of parallel classical trajectories 
hits these singular corners it splits discontinuously into two different 
bunches whose boundaries are called optical boundaries.
Quantum mechanics smoothes such singularities and associates with 
them scattering amplitudes \cite{Sommerfeld} which due to discontinuous
splitting of classical trajectories have different asymptotics at large
distances in different regions bounded by optical boundaries. Especially
complicated  case corresponds to  multiple scattering when optical boundaries
for different centers are close to each other.  
Up to now, due to singular character of such diffraction, it has only been 
proved analytically  that  the two-point correlation form factor 
for certain PI models takes at the origin  a finite value different from 
standard statistics \cite{Giraud}.

For PI systems classical periodic orbits form continuous families of parallel
trajectories restricted by singular corners. Hence, waves traveling in such
periodic orbit channels (POC) are influenced by infinite periodic arrays of 
singular diffractive centers. Fortunately, for the scattering on a staggered 
periodic array of half-planes  this problem has an exact solution found in 
\cite{Carlson} by the Wiener-Hopf method. In \cite{BS} this solution was 
analyzed in the
semiclassical limit of large energy and it was found that in the most
singular case when the incidence angle (with respect to a plane
formed by half-plane ends) is going to zero, all transmission and 
reflection coefficients tends also to zero except the `elastic' reflection
coefficient (corresponding to the specular reflection from this plane) 
which goes to $-1$. It means that the fictitious scattering plane passing
through singular diffraction corners plays 
the role of perfect mirror with the Dirichlet boundary conditions.  This 
mirror does not really exist but the multiple scattering on infinite 
number of parallel half-planes is equivalent to the reflection on this
mirror plus corrections given by complicated formulas (see
\cite{BS}) and governed by the perturbation parameter
\begin{equation}
u=\sqrt{kl}\varphi\;,  
\label{perturbation}
\end{equation}
where $\varphi$ is the incidence angle with respect to the scattering plane,
$l$ is the distance between singular corners along the scattering plane, and
$k=\sqrt{E}$ is wave momentum.
Hence, when $u\to 0$ (and $k\to\infty$) the dominant approximation to the
discussed multiple scattering problem consists in treating scattering planes
as true mirrors on which the total wave tends to zero. 

After unfolding each periodic orbit family in PPB  corresponds to an infinite 
POC restricted from both sides by straight lines passing through singular 
corners called singular diagonals (SD). When a wave with a small $u$  
moves inside such channel, it reflects back and forth from  SD
as from perfect mirrors forming a propagating wave with zero boundary 
conditions on  SD analogous to the Borrmann effect for scattering in
crystals \cite{Cole}. Therefore, in each POC one can construct the following 
quasi-states called  unfolded scar states 
which (i) obey the Dirichlet boundary conditions on SD and (ii) are periodic
(or anti-periodic) along POC
\begin{equation}
\Psi^{\mbox{\scriptsize (scars)}}_{m,n}(\xi,\eta)\sim
\sin \left (\frac{\pi}{l} m \xi +\delta \right )
\sin\left (\frac{\pi}{w}n\eta\right )\chi(\eta) \;.
\label{scar}
\end{equation}
Here $\xi$ and $\eta$ ($0<\eta<w$) are coordinates, respectively, along and
perpendicular to POC, $l$ is the length of POC equal to the length of primitive
periodic trajectory, $w$ is the channel width, $m$, $n\geq 1$ are
integers, and $\delta$ is a phase related with the  choice of
coordinates. $\chi(x)$ in (\ref{scar}) is the characteristic function of POC
($\chi(x)=1$ or $0$  when $x$ is, respectively, inside or outside POC) 
introduced to stress that scars states exist only inside POC. 
The bulk energy of such scar state is
\begin{equation}
E_{m,n}=\left (\frac{\pi}{l}\right )^2m^2+\left (\frac{\pi}{w}\right )^2n^2\;.
\label{energy}
\end{equation}
Folding back the scar state (\ref{scar}) leads to a complicated expression
$\Psi^{\mbox{\scriptsize (scar)}}_{m,n}(x,y)$ which can be 
represented in a suitable expansion basis for any system. Folded scar states
(i) obey the correct boundary conditions on billiard boundaries and (ii) 
fulfill the equation\\
$(\Delta+E_{m,n})\Psi^{\mbox{\scriptsize (scar)}}_{m,n}(x,y)=0$ everywhere
except on SD.    

The above scar states exist if the perturbation parameter 
(\ref{perturbation}) is small. As $\varphi\approx \pi n/wk$ the criterion 
of existence of a strong scar state with energy (\ref{energy})   is 
\begin{equation}
1\leq n \leq n_{\mbox{\scriptsize max}}\;\;\mbox{ and }\;\;
n_{\mbox{\scriptsize max}}\sim w\sqrt{k/l}\;.  
\label{criterion}
\end{equation}
This inequality implies that any POC in PPB will support scar states 
with fixed $n$ when $k\to \infty$ in marked contrast with the scaring on
unstable periodic orbits where contributions from individual orbits tend to
zero in semiclassical limit.

The total number of scar states depends on system considered. Only
for special type of PI models called Veech polygons \cite{Veech} (see also
\cite{BGS}) analytical  calculations are possible. For such systems  
(i) the number of POC  with $l<L$ has the quadratic asymptotics
\begin{equation}
N(l<L)\stackrel{L\to\infty}{\longrightarrow} CL^2/A
\label{Np}
\end{equation}
where $A$ is the billiard area, $C$ is  a system dependant constant, 
and (ii)  the width of POC with length  $l$  is 
\begin{equation}
w=\gamma A/l
\label{wp}
\end{equation}
where $\gamma<1$ is a constant taken from a finite set.  

\begin{figure}[ht]
\begin{center}  
\epsfig{file=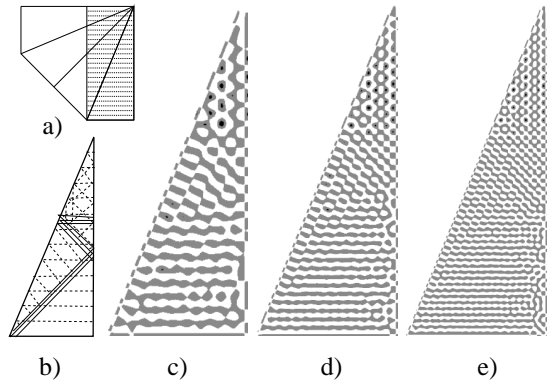, angle=270, width=\figwidth} 
\end{center}

\vspace{-1ex}

\caption{
  {\bf a)} Unfolded scar state for the simplest POC of the right triangle
  with angle $\pi/8$. {\bf b)} Schematic folding of this state.
  Dashed lines indicate its maxima.
  Three solid lines show a region near SD where unfolded scar function
  tends to zero. {\bf c)}-{\bf e)} Eigenfunctions with energy $E$ close to
  the scar energy $E_{m,n}$.
  {\bf c)} $E=407.4$, $E_{50,1}=407.6$. 
  {\bf d)} $E=1015.97$, $E_{79,1}=1016.12$.
  {\bf e)} $E=1968.97$, $E_{110,1}=1969.15$.  }
\label{fold}
\end{figure}

Using (\ref{criterion}) and these formulas one concludes that  (i) scar states
exist only for POC whose length is restricted
\begin{equation}
l\leq l_{\mbox{\scriptsize max}}\;\; \mbox{ with }\;\;
l_{\mbox{\scriptsize max}} \sim k^{1/3}
\end{equation}
(other channels are closed and
cannot support propagating waves with $n\geq 1$) and (ii) the averaged 
density of scar states $\bar{\rho}_s$ is of the same order as the mean total 
density of states $\bar{\rho}$
\begin{equation}
\bar{\rho}_s=\sum_{\mbox{\scriptsize scars}}
\delta(E-E_{m,n})\sim \sum_{l<l_{\mbox{\scriptsize max}}}
\frac{l}{k}n_{\mbox{\scriptsize max}}\sim \bar{\rho}\;.
\label{srho}
\end{equation}
These results show that for PPB where (\ref{Np}) and (\ref{wp}) hold
scar states are  a good zero-order approximation to wave functions.
As an illustration consider e.g. the right triangle with angle $\pi/8$.
Its simplest POC corresponds to orbits perpendicular to the shortest side of
the triangle. After unfolding it has a rectangular 
shape indicated in  Fig.~\ref{fold}a. The folded scar state 
for this POC is shown schematically in Fig.~\ref{fold}b. Dashed lines in 
this figure  indicate maxima of the scar state. They have
complicated form except at the right corner of the triangle where they form
horizontal lines. 

In Figs.~\ref{fold}c-\ref{fold}d three true eigenfunctions of this
triangular billiard (with area $4\pi$) are presented. The eigenfunctions
were chosen because their energies are close to the scar energy
(\ref{energy}) calculated with $l=a$ and $w=b$ where $a$ and $b=a\tan \pi/8$ 
are sides of the triangle.
The characteristic horizontal lines corresponding to the scar picture (as in
Fig.~\ref{fold}b) are clearly seen in all these eigenfunctions.

\vspace{-1ex}

\noindent
These pictures are just a few examples (amongst many others) of clear scar
eigenfunctions observed in triangular billiards.
Complicated folding of POCs in such models makes it difficult to
visualize scar states associated with longer trajectories. This goal can more
easily be achieved in another PI model, called barrier billiard (BB), which in
the simplest case consists of a rectangular billiard with the Dirichlet
boundary conditions on all sides except the half of one side on which 
the Neumann boundary condition  is imposed \cite{Wiersig}. In this model POCs
are the same as for integrable rectangular billiard and are specified by two
coprime integers $M$, $N$. The POC length is $l=\sqrt{(2aM)^2+(2bN)^2}$
where $a,\;b$ are sides of the rectangle. The usual
POCs for rectangular billiard are splitted (and restricted) by images of the
singular point. For odd $M$  each POC is divided into two POCs
of width $w=2ab/l$. Both channels can support scar states but one
requires odd $m$  and the other - even $m$. When $M$ is even, POC remains
unramified and has the width $w=4ab/l$.

\begin{figure}[ht]
\begin{center}  
\epsfig{file=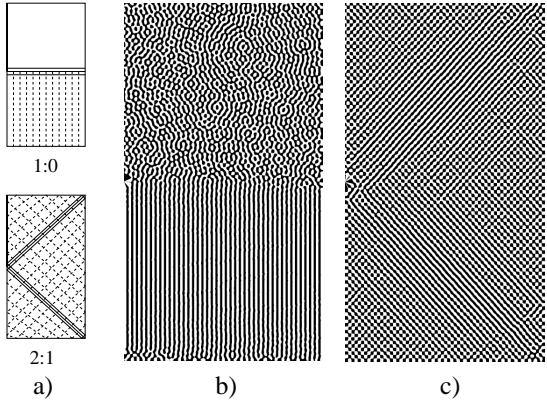, angle=270, width=\figwidth}  
\end{center}

\vspace{-2ex}

\caption{BB eigenfunctions  strongly influenced by
  1:0 and 2:1 scar states. {\bf a)} Folded scar states for the 1:0 (top) and
  2:1 (bottom) periodic trajectories.  Thick line shows the part of BB 
  with the Neumann  boundary condition. Dashed lines indicate  maxima of
  the scar state. Three solid lines show regions around SD where 
  unfolded scar states tend to zero.  {\bf b)} Eigenfunction with 
  $E=10209.55$. The 1:1 scar energy
  $E_{85,1}=10209.65$. {\bf c)} Eigenfunction with  
  $E=10017.57$. The 2:1 scar energy $E_{453,1}=10017.67$.    }
\label{bball}
\end{figure}

\begin{figure}[ht]
\begin{center}  
\epsfig{file=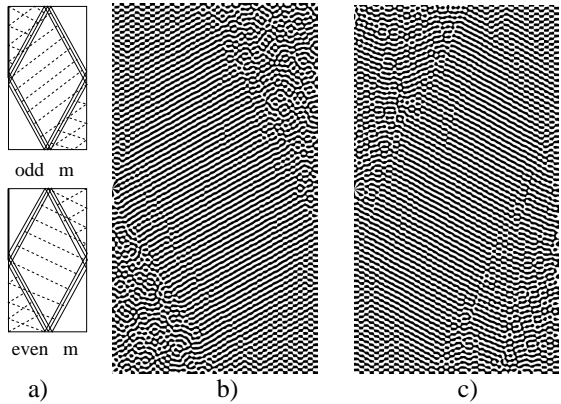, angle=270, width=\figwidth}  
\end{center}

\vspace{-2ex}

\caption{The same as in Fig.~\ref{bball} but for  1:1 scar states. 
  {\bf a)} Folded scar states for  1:1 POCs
  with even and odd $m$.
  {\bf b)} Eigenfunction with $E=10041.41$. The 1:1 scar energy
  $E_{347,1}=10041.87$. {\bf c)} Eigenfunction
  with $E=10099.58$.  The 1:1 scar energy $E_{348,1}=10099.82$.    }
\label{left_right}
\end{figure}

In Figs.~\ref{bball} and \ref{left_right} a few examples of high-excited
scar
\vspace{-2ex}

\vspace{-5ex} 

\noindent
states for BB with $(b/a)^2=\sqrt{5}+1$ and $ab=4\pi$ are presented.
Black and white regions in these figures correspond to  positive and negative
values of eigenfunctions which  are small in regions with irregular nodal 
patterns. This nodal domain representation is quite sensitive 
because even a week noise changes drastically regular nodal pictures.
Nevertheless, these (and many other) pictures show high-quality scar 
structures for BB. Fixing a POC and increasing the energy we always find
reasonably good picture of the corresponding scar state close to the scar 
energy (\ref{energy}). 

\begin{figure}[ht]
\begin{center}  
\epsfig{file=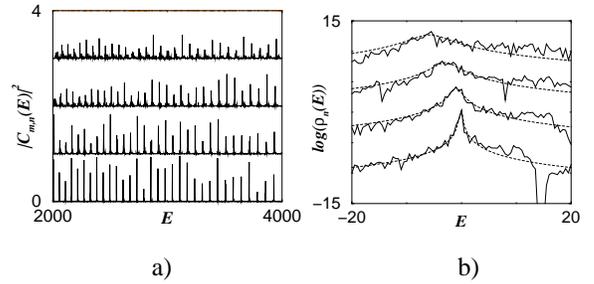, angle=270, width=\figwidth}  
\end{center}

\vspace{-6ex}

\caption{{\bf a)} Overlap of exact eigenfunctions of BB with the 1:1 scar
  state with (from bottom to top) $n=1,2,3,4$ (graphs for different
  $n$ are shifted up for clarity by $n-1$ units).
 {\bf b)} Local density   (\ref{local}) for this overlap
  (graphs with different $n$ are shifted up by $5(n-1)$ units).
  Dashed lines indicate the best fit in the Breit-Wigner form (\ref{BW}). }
\label{overlap}
\end{figure}

More quantitative description of scar states is achi\-e\-ved by computing the
overlap of folded scar states $\Psi^{\mbox{\scriptsize (scar)}}_{m,n}(x,y)$ 
with exact eigenfunctions $\Psi_{E_{\lambda}}(x,y)$
\begin{equation}
C_{m,n}(E_{\lambda})=\int 
\Psi^{\mbox{\scriptsize (scar)}}_{m,n}(x,y)\Psi_{E_{\lambda}}(x,y)dxdy\;. 
\end{equation}
In computations we fix $n$ and choose  
$m$ from the condition of  minimum of  $|E_{\lambda}-E_{m,n}|$
(when $m$ is kept fixed only one peak appears).  In Fig.~\ref{overlap}a
we plot $|C_{m,n}(E)|^2$ versus $E$ with $2000<E_{\lambda}<4000$  
for the 1:1 scar state.

\vspace{-6ex}

\noindent
Spikes in such  figures confirm that practically near all 
scar energies (\ref{energy}) there exists true eigenfunctions which have  
strong contribution from the given scar state.  
The analysis of these spikes reveals that their local density 
\begin{equation}
\rho_n(E)=\left \langle \sum_{\lambda} 
|C_{m,n}(E_{\lambda})|^2 \delta(E-E_{\lambda}+E_{m,n})\right \rangle_{m}  
\label{local}
\end{equation}
averaged over different $m$ can well be approximated by the Breit-Wigner 
distribution (see Fig.~\ref{overlap}b) 
\begin{equation}
\rho_n(E)=\frac{\Gamma_{n}(E)}{2\pi
  [(E-\epsilon_{n}(E))^2+\Gamma_{n}^2(E)/4]}  
\label{BW}
\end{equation}
similar to the one observed in random band matrices with
preferential basis \cite{Shepelyansky}. For the 1:1 scar state the best fit
gives $\Gamma_n\approx 3.5n^2/\sqrt{k}$ which agrees qualitatively with 
an estimate which can be obtained from  \cite{BS} that for BB the total width 
of a given scar state in the leading order  is   
$\Gamma_{n}(E) \sim  (n^2/w^2)\sqrt{l/kw^2}$.

For Veech billiards the density of scar states is of the same
order as the total density of states (cf. (\ref{srho})) and one is lead to 
to the conjecture that their eigenfunctions can be represented as a sum over 
scar states
\begin{equation}
\Psi_{E_{\lambda}}(x,y)=\sum_{\mbox{\scriptsize scars}}
C_{m,n}(E_{\lambda})\Psi_{m,n}^{\mbox{\scriptsize (scars)}}(x,y)\;.
\label{statistical}
\end{equation}
For chaotic systems ergodicity according to Shnirelman's theorem
\cite{Shnirelman} forces scar states to be rare. PI models are not ergodic 
and no a-priori objections exist which  prevent this type of superscaring.  

Important characteristic of PPB wave functions is a set of participation
ratios (see e.g. \cite{Chalker}, \cite{Mirlin} and references therein) 
\begin{equation}
R_q(E)= (\sum_{m,n}|A_{m,n}(E)|^{2q} )^{-1}
\end{equation}
where $A_{m,n}(E)$ are coefficients of the expansion of wa\-ve functions in a
suitable basis normalized such that $\sum_{m,n}|A_{m,n}|^2=1$. 
PI systems, as all systems with intermediate statistics, should have some 
fractal properties \cite{Chalker} and it is naturally to expect that
$R_q(E)\stackrel{k\to\infty}{\longrightarrow} k^{D_q(q-1)}$ with fixed $D_q$ 
called generalized fractal dimensions (see e.g. \cite{Chalker}, \cite{Mirlin}).

\begin{figure}[h]
\begin{center}  
\epsfig{file=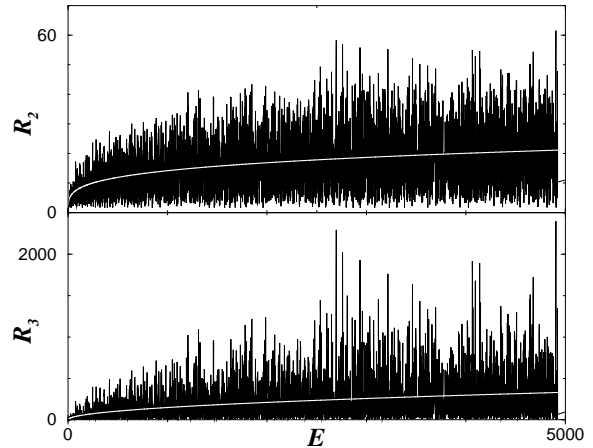, angle=270, width=\figwidth}  
\end{center}

\vspace{-2ex}

\caption{Participation ratios $R_2$ (top) and $R_3$ (bottom) versus energy 
  for BB. White lines indicate the fits  $R_2\approx 2.52\sqrt{k}$ and
   $R_3\approx 4.7k$.   
   }
\label{participation}
\end{figure}

Under the simplest assumption that for scar states the local density of
$|C_{m,n}(E)|^{2q}$ is proportional to $\rho_n^q(E)$
with the Breit-Wigner form (\ref{BW}) of $\rho_n(E)$, 
(\ref{statistical}) and (\ref{BW}) with the above
estimate for $\Gamma_n$  give  $D_q=.5$ confirming 
fractal character of BB eigenfunctions in momentum representation. 
In Fig.~\ref{participation} we plot $R_q$ for $q=2$ and $q=3$ for BB computed
directly from the expansion of eigenfunctions into trigonometric series.
The best fits $R_2\approx 2.52\sqrt{k}$ and $R_3\approx 4.7k$ very well
describe the data in the given interval of energy which means that for BB
$D_2\approx D_3\approx .5$ in accordance with the above estimates. 
The assumed value of spectral compressibility for BB, $\chi=.5$, 
\cite{Wiersig} is close to the spectral compressibility $\chi(D_2)$  
numerically computed at the point $D_2=.5$ for the critical power-law 
random band matrix model (cf. Fig. 2 of \cite{Mirlin}).

In summary, we have argued that strong diffraction in PPB leads 
to the formation of new type of long-lived resonant states (scar states) 
propagating inside POC and reflecting from SD as from perfect mirrors. Many 
true eigenfunctions  of PPB have surprisingly clear structures associated with 
such states even at high energies. It follows  from our results that PPB 
are the best models of scar phenomenon.  
For good PI models (Veech billiards) the  density of scar states is a constant
and they can be considered as the basis of perturbation expansion. 
A week residual  interaction between them (neglected in this paper) forms 
true eigenstates and leads to intermediate character of spectral statistics 
for these models. It appears that this interaction shares many features with 
critical random band matrix model. We have also checked that BB wave functions
have fractal properties in momentum space.

\vspace{-2ex}

\begin{thebibliography}{99}
\bibitem[\dag]{ad} Unit\'e de recherche de l'Universit\'e de Paris XI 
  associ\'ee au CNRS.
\bibitem{BGS} O. Bohigas, M.-J. Giannoni, and C. Schmit, Phys. Rev. Lett. 
  {\bf 52}, 1 (1984).
\bibitem{Berry} M. V. Berry, J. Phys. A: Math. Gen. {\bf 10},  2083 (1977).
\bibitem{Richens} P. J. Richens and M.V. Berry, Physica D {\bf 2}, 495 (1981).
\bibitem{Gerland} E. B. Bogomolny, U. Gerland, and C. Schmit, Phys. Rev. E
  {\bf 59}, R1315 (1999).
\bibitem{Shklovskii} B. I. Shklovskii {\it et al.}, Phys. Rev. B {\bf 47},
  11487  (1993).
\bibitem{Heller1} E.J. Heller, Phys. Rev. Lett. {\bf 53}, 1515 (1984).
\bibitem{Heller2} L. Kaplan and R.J. Heller, Ann. Phys. {\bf 264}, 171 (1998).
\bibitem{Sommerfeld} A. Sommerfeld, {\it Optics} (New York, Academic, 1954). 
\bibitem{Giraud} E. Bogomolny, O. Giraud, and C. Schmit, Commun. Math. Phys.
  {\bf 222}, 327 (2001).     
\bibitem{Carlson} J.F. Carlson and A.E. Heins, Quart. Appl. Math. {\bf 4},
  313 (1947).
\bibitem{BS} E. Bogomolny and C. Schmit, Nonlin. {\bf 16}, 2035 (2003).  
\bibitem{Cole} B.W. Batterman and H. Cole, Rev. Mod. Phys. {\bf 36}, 681 
  (1964).
\bibitem{Veech} W.A. Veech, Invent. Math. {\bf 97}, 533  (1989).
\bibitem{Wiersig} J. Wiersig, Phys. Rev. {\bf 65},  046217 (2002).
\bibitem{Shepelyansky} Ph. Jacquod and D.L. Shepelyansky, Phys. Rev. Lett.
  {\bf 75},  3501 (1995).
\bibitem{Shnirelman} A.I. Shnirelman, Usp. Math. Nauk. {\bf 29}, 181  (1974).
\bibitem{Chalker} J.T. Chalker, V.E. Kravtsov, and I.V. Lerner, JETP Lett.
  {\bf 64},  386 (1996).
\bibitem{Mirlin} F. Evers and A.D. Mirlin, Phys. Rev. Lett. {\bf 84},
  3690  (2000). 
\end{thebibliography}
\end{document}